# Does Moiré Matter? Critical Moiré Dependence with Quantum Fluctuations in Graphene Based Integer and Fractional Chern Insulators

**Authors:** Zihao Huo[1]*, Wenxuan Wang[1]*, Jian Xie[1]*, Yves H. Kwan[2]*, Jonah Herzog-Arbeitman[2], Zaizhe Zhang[1], Qiu Yang[1], Min Wu[1], Kenji Watanabe[3], Takashi Taniguchi[4], Kaihui Liu[5], Nicolas Regnault[2,6,7], B. Andrei Bernevig[2,8,9]†, and Xiaobo Lu[1,10]†

**Affiliations:**

[1]International Center for Quantum Materials, School of Physics, Peking University, Beijing 100871, China

[2]Department of Physics, Princeton University, Princeton, New Jersey 08544, USA

[3]Research Center for Electronic and Optical Materials, National Institute for Material Sciences, 1-1 Namiki, Tsukuba 305-0044, Japan.

[4]Research Center for Materials Nanoarchitectonics, National Institute for Material Sciences, 1-1 Namiki, Tsukuba 305-0044, Japan

[5]State Key Laboratory for Mesoscopic Physics, Frontiers Science Centre for Nano-optoelectronics, School of Physics, Peking University, Beijing 100871, China

[6]Center for Computational Quantum Physics, Flatiron Institute, 162 5$^{th}$ Avenue, New York, NY 10010, USA

[7]Laboratoire de Physique de l'Ecole Normale Supérieure, ENS, Université PSL, CNRS, Sorbonne Université, Université Paris-Diderot, Sorbonne Paris Cité, Paris, France

[8]Donostia International Physics Center, P. Manuel de Lardizabal 4, 20018 Donostia-San Sebastian, Spain

[9]IKERBASQUE, Basque Foundation for Science, Bilbao, Spain

[10]Collaborative Innovation Center of Quantum Matter, Beijing 100871, China

*These authors contributed equally to this work.

†E-mail: bernevig@princeton.edu; xiaobolu@pku.edu.cn;

**Abstract:**

Rhombohedral multilayer graphene has emerged as a powerful platform for investigating flat-band-driven correlated phenomena, yet most aspects remain not understood. In this work, we systematically study the moiré-dependent band topology in rhombohedral hexalayer graphene. For the first time we demonstrate that the moiré twist angle plays a crucial role in the formation of the moiré Chern insulators in rhombohedral hexalayer graphene/hexagonal boron nitride (RHG/hBN) moiré superlattices. In the moiré-distant regime at filling factor $v = 1$, only systems with a twist angle $\theta < 1.1°$ exhibit an integer moiré Chern insulator, while the fractional Chern insulator at $v = 2/3$ requires smaller twist angle to be stabilized. Our theoretical modelling, which includes quantum fluctuations and exact diagonalization results, suggests that mean-field theory, which has been widely adopted, does not explain the twist-angle dependence of the $v = 1$ phase diagram, and that correlation effects are crucial. Moreover, we realize two distinct stacking configurations ($\xi = 0$ and $\xi = 1$) between graphene and hBN, and find that both cases can yield a Chern insulator at $v = 1$. Our experimental work upends the current mean-field paradigm, illuminates how quantum fluctuations and moiré effects shape the RHG/hBN phase diagram, and paves the way for future understanding and engineering of topological correlated states in rhombohedral graphene moiré systems.

# Main Text:

The experimental observation of fractional Chern insulators (FCIs) in rhombohedral multilayer graphene aligned with hBN[1–6] has established this material family as the second experimental system after twisted MoTe$_2$[7–16] for possible studies of charge fractionalization[7–11] and (so far) Abelian anyonic statistics[12–15] at zero magnetic field. FCIs have been observed in the moiré-distant regime where an external displacement field pushes the doped electrons away from the aligned hBN, when the rhombohedral graphene consists of n=4,5,6 layers[1–6]. Similar to twisted MoTe$_2$[16–19], the formation of such FCIs in rhombohedral graphene systems is accompanied by spin-valley polarized Chern insulators at integer filling $v = 1$ (one electron per moiré unit cell). Experimentally, such integer Chern insulators (ICIs) have been widely observed across different rhombohedral graphene systems ($3 \leq n \leq 8$)[1–4,20–24]. In twisted MoTe$_2$, it has been demonstrated that the twist angle has significant impacts on the band topology[16–19,25–32]. However, the twist angle ($\theta$) dependence of the band topology in the rhombohedral graphene/hBN superlattices remains uncharted experimentally. Theoretically, the effect of the twist angle on band topology in rhombohedral graphene/hBN superlattices is challenging to discern due to the nearly gapless nature of the low-energy moiré bands, making them highly susceptible to interactions[33–40] which may not be adequately captured by mean-field theory[41]. Moreover, despite sharing the same moiré periodicity, two microscopically distinct stacking configurations ($\xi = 0$ and $\xi = 1$) originating from the broken C$_{2z}$ symmetry in both hBN and rhombohedral graphene create fundamentally inequivalent moiré structural isomers[42]. The effect of the stacking configuration on band topology needs to be unravelled.

In this work, we present a multi-dimensional demonstration of the unconventional moiré dependence of Chern insulators in RHG/hBN superlattices. We address how the two stacking configurations ($\xi = 0$ and $\xi = 1$) with similar twist angles affect the electronic topology, how different twist angles affect both ICIs and FCIs, and how the correlated states evolve when the electrons are pushed close to, or away, from the moiré interface. While doing so, we remarkably find experimental results that run opposite to Hartree-Fock (HF) predictions[33–37], but are in better agreement with beyond mean-field and exact diagonalization (ED) results and/or moiré-induced correlated states.

**Moiré Chern insulator in RHG/hBN with different stacking configurations**

Since both rhombohedral multilayer graphene and hBN break C$_{2z}$ symmetry, the same moiré periodicity may correspond to different stacking configurations, as illustrated in Fig. 1a. Stacking geometry dependent band topology has been observed in MoTe$_2$/WSe$_2$ heterostructures[43,44]. So far, most theoretical studies of rhombohedral multilayer graphene/hBN systems have not investigated the impact of the stacking. In the moiré-distant regime at $v = 1$, previous HF calculations[37] predict that the $\xi = 0$ and $\xi = 1$ configurations exhibit quantitative differences in the phase diagram of rhombohedral graphene/hBN. While the doped electrons are driven away from the aligned hBN, it has been suggested that the Chern number $C = 0$ or 1 of the interacting ground state is nevertheless influenced by the stacking-dependent moiré properties of the occupied valence bands. Beyond HF calculations, it is unclear what, if any, the effect of the stacking is, considering current extensive exact diagonalization studies[41] do not yet show topological states of matter, especially at FCI fillings.

However, experimentally distinguishing the two stacking configurations, in particular identifying the lattice details of rhombohedral multilayer graphene, remains challenging. As an alternative, we utilized monolayer hBN surface steps to realize both stacking structures within the same sample. As shown in Fig. S1a, the hBN lattice undergoes an effective 180° relative rotation on either side of a monolayer hBN step. We chose a top-surface monolayer-stepped hBN as the bottom hBN (Fig. S1b), where RHG flakes with the same orientation were patterned on both sides of the step and aligned, enabling simultaneous 0° and 180° alignment between the RHG and hBN. Figs. 1b-c show the second-harmonic generation (SHG) characterization of the hBN verifying the alternating odd/even layer configuration across the monolayer step. In sharp contrast with odd-layer hBN, the SHG signal vanishes in even-layer hBN due to the presence of inversion symmetry. Further evidence demonstrating the monolayer step of hBN proximate to RHG is given by the atomic force microscopy measurement (Methods). Fig. 1d presents an optical microscope image of the fabricated devices, D3 and D5. The region indicated by the red dashed box corresponds to one stacking configuration $\xi = 0$ or $\xi = 1$, while the blue region corresponds to the other configuration. Both regions exhibit large moiré periods, with $\theta = 0.42°$ and $\theta = 0.63°$ (Methods). Fig. 1e (f) shows the variation of $\rho_{xx}$ ($\rho_{xy}$) in device D3 with one stacking geometry and $\theta = 0.42°$ as a function of moiré filling factor $v$ and displacement field $D/\varepsilon_0$, with a local minimum (maximum) observed in $\rho_{xx}$ ($\rho_{xy}$) at $v = 1$. Similar behavior is observed in device D5 with the other stacking geometry and $\theta = 0.63°$ (Figs. 1j-k). Moreover, Figs. 1g-h show the Landau fan diagrams of device D3, revealing a distinctive Chern gap with $C = 1$ which persists to zero magnetic field. This topological gap exhibits exceptional stability compared to Landau level gaps, which typically vanish below a critical external field. Meanwhile, the $v = 1$ state in this moiré-distant regime exhibits clear quantum anomalous Hall effect with vanishing $\rho_{xx}$ and more than 95% quantized $\rho_{xy}$ (Fig. 1i). Similar behavior is observed in device D5 as shown in Figs. 1l-n, demonstrating that the $C = 1$ Chern insulator can be stabilized in both stacking geometries. The discrepancy between our experimental observations and previous HF theoretical predictions[37] underscores the necessity for developing theories that incorporate effects beyond the mean-field approximation.

**Moiré dependence of band topology**

In order to demonstrate the moiré twist angle dependence of band topology clearly, we fabricated a series of devices with varying twist angles. Measurements on device D3 and D5 reveal that, unlike the correlated insulator states at the moiré-proximal side[42], the $v = 1$ state with $C = 1$ at the moiré-distant side remains robust against the two stacking configurations. In Figs. 2a-g, we show $\rho_{xx}$ and $\rho_{xy}$ data of seven devices with increasing twist angles, presenting the evolution of the $v = 1$ state in the moiré-distant regime. By combining the data from Figs. 2a-g, we chart both $\rho_{xx}$ and $\rho_{xy}$ linecuts (Fig. 2h) as a function of $D/\varepsilon_0$ taken in the $v = 1$ state, as well as a schematic phase diagram (Fig. 4c) of the $C = 1$ state at $v = 1$ as a function of twist angle, tracking the critical displacement field $D_c$. Our measurements reveal that the $C = 1$ state persists only in devices with small twist angles ($\theta < 1.1°$). With increasing twist angle, the critical displacement field $D_c$ required to stabilize the $C = 1$ state shifts to higher values, while its range narrows, until the $C = 1$ state eventually collapses by $\theta = 1.10°$ at zero magnetic field.

We have also systematically investigated the topological properties at fractional moiré fillings in devices D4, D6, D7, presenting both their phase diagrams (Figs. 3a-c) and Landau fan diagrams (Figs. 3d-f) at optimal displacement field. At relatively small angle, D1 ($\theta = 0.17°$) and D2 ($\theta =$

0.22°) exhibit both FCIs and ICIs (more details shown in previous work[2]). D3 and D5 yield unreliable data because of poor contacts at the fractional filling regions. In device D4 with $\theta$ = 0.49°, the $\rho_{xx}$ exhibits a distinct minimum at $v$ = 2/3 (Fig. 3a). Both $\rho_{xx}$ and $\rho_{xy}$ disperse with magnetic field and carrier density following Streda's formula ($\partial n/\partial |B_\perp| = C \cdot e/h$) with $C$ = 2/3 (Fig. 3d), showing signatures of FCI state at $v$ = 2/3 with $C$ = 2/3. This FCI state is also accompanied by the sign reversal of $C$ at high magnetic fields, consistent with the observation of our previous work[2]. On the other hand, in device D6 with $\theta$ = 0.92°, a Chern insulator with $C$ = 1 originating from $v$ = 2/3 (Figs. 3b,e) is observed, instead of $C$ = 2/3. However, by the time the angle increases to $\theta$ = 1.10° in device D7 (Figs. 3c,f), no topologically non-trivial state can be resolved at $v$ = 2/3 from the Landau fan diagram.

Our results demonstrate that a small twist angle favors the formation of both fractional and integer Chern insulators in RHG/hBN at zero magnetic field. Compared to ICIs, FCIs show stronger angular sensitivity. We anticipate that similar moiré dependent behaviors can be expected in pentalayer and tetralayer systems[1,3,5,24,45]. Our experiments also reveal a moiré-dependent symmetry broken state at $v$ = 0 and $D$ = 0 which has also been observed in tetralayer and pentalayer systems[46,47]. As shown in Fig. S2, the symmetry broken state can be significantly suppressed by the moiré potential at small angle, while it is robust in the devices with larger twist angles. We also note that the data of D5 with $\theta$ = 0.63° show a slightly weaker state symmetry broken state at $v$ = 0 and $D$ = 0 than data shown in D3 with $\theta$ = 0.42° and D4 with $\theta$ = 0.49°. Such a deviation from the overall trend might be from the two stacking configurations as discussed in the previous section.

Note that similar to pentalayer and tetralayer systems[1,3,5,24], zero-field FCIs and ICIs of RHG system are stabilized away from moiré interface corresponding to $D$ < 0. Here, we also investigate the behaviors of electrons close to the moiré interface with $D$ > 0 (Fig. S3,4). Instead of forming ICIs or FCIs at these fillings, a trivial insulator at $v$ = 1 and trivial charge density waves at fractional fillings ($v$ = 1/3 and 2/3, Fig. S4a-b) are more favorable. Fig. S3e displays $\rho_{xx}$ versus filling $v$ taken from the positions indicated with the dashed lines in Fig. S3a,4d, showing resistive peaks at $v$ = 1/3, 2/3 and 1. This is consistent with non-interacting calculations[48] showing that the conduction band is trivial on the moire-proximate side at zero magnetic field.

Interestingly, with finite magnetic field $B_\perp$ applied to the device with small moiré twist angle ($\theta$ = 0.22°), both $v$ = 1 and 2/3 states exhibit clear $B_\perp$-dependence which can be well described by the Streda's formula with $C$ = -1 (Fig. S3b,4b). Moreover, similar $B_\perp$-dependence also emerges at $v$ = 1/2 (Fig. S3c,4a). The observation of such magnetic field dependence at $v$ = 2/3 and 1/2 indicates the formation of symmetry-broken ICIs with tripling and doubling of the original moiré unit cell respectively. Fig. S3d displays the $B_\perp$-dependence of $\rho_{xx}$ and $\rho_{xy}$ along the dashed lines in Fig. S3b-c. Compared with the $C$ = -1 state at $v$ = 1, the $B_\perp$-stabilized $C$ = -1 states at $v$ =1/2 and 2/3 are more fragile. At high magnetic field, the $C$ = -1 states at $v$ = 1/2 and 2/3 disappear. Similar behaviors have been observed by measuring the electronic compressibility in pentalayer graphene/hBN[4]. Increasing or decreasing the magnetic field would close the gap at these fractional fillings.

**The importance of quantum fluctuations**

The observed dependence of the Chern number in the moiré-distant regime at $v = 1$ on the moiré twist angle (Fig. 4c) is intriguing, since it runs completely opposite to all current mean-field predictions[33–37]. As shown in Fig. 4d, HF calculations on an interacting continuum model of RHG/hBN, which incorporates the electrostatic moiré potential of the filled valence bands[37,49], find that the topologically trivial $C = 0$ state has a lower energy than the Chern insulator for smaller twist angles, and vice versa for larger twist angles. This behavior is reproduced for other choices of single-particle and interaction parameters (see Supplementary Information). This HF trend can be motivated from the artificial limit of vanishing moiré potential, where the twist angle simply sets the size of the moiré Brillouin zone (BZ). Theoretical mean-field studies[39,50–52] have shown that lower twist angles relatively favor the $C = 0$ state, because the moiré BZ encloses a smaller Berry curvature flux.

A candidate explanation for the stark disagreement between HF calculations and experimental observations is that mean-field theory is inadequate for correctly resolving the phase diagram at $v = 1$[41]. Fig. 4d shows that for most twist angles, the $C = 0,1$ states in HF are separated by small energy differences (< 0.1meV per unit cell), which may be overpowered by quantum fluctuations beyond mean-field theory (schematized in Figs. 4a-b). To investigate this possibility, we perform calculations on the same interacting continuum model using two beyond-HF methods (Fig. 4e): "bandmax"-truncated exact diagonalization (ED)[37,53] and the generalized random phase approximation (gRPA)[54,55]. Both techniques introduce quantum fluctuations on top of the uncorrelated $C = 0,1$ HF wave functions, and compute the (negative) correlation energy $E^{corr} = E - E^{HF}$, which measures the lowering of the total energy $E$ compared to the mean-field result $E^{HF}$. Fig. 4f plots the correlation energy in ED for different truncation parameters $\{N, 0\}$, where larger $N$ means that more fluctuations are included in the Hilbert space (Methods). We observe that $E^{corr}$ for the $C = 1$ state becomes increasingly more negative relative to the $C = 0$ state for lower twist angles. We also obtain the same qualitative dependence on $\theta$ in the gRPA calculation shown in Fig. 4g. The magnitudes of the correlation energies are significantly larger than in the ED calculations, which suggests a tendency of gRPA to over-correct the mean field results.

Our theoretical results demonstrate that correlation effects are acting to reverse the twist-angle dependence of the mean-field phase diagram and possibly bridge the discrepancy with experiments. Fig. 4g shows that the gRPA calculation predicts a full reversal where the $C = 1$ ($C = 0$) state is stabilized at the lower (higher) end of the twist angle range $\theta \in [0.0° - 0.8°]$. This would *seem* in agreement with the experiments, but the gRPA likely overestimates the correlation energy. The ED calculation exhibits a similar inversion, though the ground state remains $C = 0$ for the smallest twist angles $\theta < 0.2°$, and we caution that the calculation is limited to small system sizes and truncation parameters. In the Supplementary Information, we examine alternative choices of single-particle parameters and interaction schemes, where gRPA and ED calculations are not always able to reverse the HF phase diagram. For this reason, we believe there are still missing ingredients either in the single particle Hamiltonian or in the interaction scheme that could give rise to the robust behavior seen in the experiments. Nevertheless, we consistently find that the correlation energies relatively favor the $C = 1$ ($C = 0$) state for lower (higher) twist angles. While there are other potential factors, such as the hitherto unexplored possibility that the single-particle moiré parameters significantly vary with $\theta$, our analysis points to quantum fluctuations playing a crucial role in determining the experimental phase diagram of correlated insulators at $v = 1$.

## Conclusions

In conclusion, we have reported for the first time the change in the topological phase diagram with the moiré twist angle in RHG/hBN, proving the essential role of moiré in the correlated states. We have demonstrated that different stacking configurations in RHG/hBN can give rise to a $v = 1$ Chern insulator in the moiré-distant regime. Our novel experimental technique based on a monolayer hBN step edge can be leveraged for further investigations of the stacking dependence. Second, and most importantly, we found that both FCIs and ICIs are sensitive to the moiré twist angle, and can only be stabilized within a certain angle range. We also observed when electrons are polarized proximate to the moiré interface, the system exhibits topologically trivial integer and fractional states. With finite magnetic field applied, samples with small moiré twist angle exhibit signatures of symmetry-broken ICIs at fractional fillings ($v = 1/2$ and $2/3$). We argue that the $\theta$-dependence of the topological phase diagram at $v = 1$ cannot be explained by mean-field theory and requires inclusion of correlation effects and/or further consideration of moiré effects in the Hamiltonian. Our work establishes a global topological phase diagram for RHG/hBN superlattices and motivates analogous studies of other numbers of layers. It also provides crucial insights for effectively understanding and constructing related topological correlated states in such systems.

## References and Notes

1. Lu, Z. *et al.* Fractional quantum anomalous Hall effect in multilayer graphene. *Nature* **626**, 759–764 (2024).
2. Xie, J. *et al.* Tunable fractional Chern insulators in rhombohedral graphene superlattices. *Nat. Mater.* **24**, 1042–1048 (2025).
3. Lu, Z. *et al.* Extended quantum anomalous Hall states in graphene/hBN moiré superlattices. *Nature* **637**, 1090–1095 (2025).
4. Aronson, S. H. *et al.* Displacement Field-Controlled Fractional Chern Insulators and Charge Density Waves in a Graphene/hBN Moiré Superlattice. *Phys. Rev. X* **15**, 031026 (2025).
5. Choi, Y. *et al.* Superconductivity and quantized anomalous Hall effect in rhombohedral graphene. *Nature* **639**, 342–347 (2025).
6. Xie, J. *et al.* Unconventional Orbital Magnetism in Graphene-based Fractional Chern Insulators. Preprint at https://doi.org/10.48550/arXiv.2506.01485 (2025).
7. Regnault, N. & Bernevig, B. A. Fractional Chern Insulator. *Phys. Rev. X* **1**, 021014 (2011).
8. Neupert, T., Santos, L., Chamon, C. & Mudry, C. Fractional Quantum Hall States at Zero Magnetic Field. *Phys. Rev. Lett.* **106**, 236804 (2011).
9. Sheng, D. N., Gu, Z.-C., Sun, K. & Sheng, L. Fractional quantum Hall effect in the absence of Landau levels. *Nat. Commun.* **2**, 389 (2011).
10. Sun, K., Gu, Z., Katsura, H. & Das Sarma, S. Nearly Flatbands with Nontrivial Topology. *Phys. Rev. Lett.* **106**, 236803 (2011).
11. Tang, E., Mei, J.-W. & Wen, X.-G. High-Temperature Fractional Quantum Hall States. *Phys. Rev. Lett.* **106**, 236802 (2011).
12. Nakamura, J., Liang, S., Gardner, G. C. & Manfra, M. J. Direct observation of anyonic braiding statistics at the $v=1/3$ fractional quantum Hall state. *Nat. Phys.* **16**, 931–936 (2020).
13. Bartolomei, H. *et al.* Fractional statistics in anyon collisions. *Science* **368**, 173–177 (2020).

14. Samuelson, N. L. *et al.* Anyonic statistics and slow quasiparticle dynamics in a graphene fractional quantum Hall interferometer. Preprint at https://doi.org/10.48550/arXiv.2403.19628 (2024).
15. Werkmeister, T. *et al.* Anyon braiding and telegraph noise in a graphene interferometer. *Science* eadp5015 (2025).
16. Park, H. *et al.* Observation of fractionally quantized anomalous Hall effect. *Nature* **622**, 74–79 (2023).
17. Xu, F. *et al.* Observation of Integer and Fractional Quantum Anomalous Hall Effects in Twisted Bilayer MoTe 2. *Phys. Rev. X* **13**, 031037 (2023).
18. Anderson, E. *et al.* Programming correlated magnetic states with gate-controlled moiré geometry. *Science* **381**, 325–330 (2023).
19. Cai, J. *et al.* Signatures of fractional quantum anomalous Hall states in twisted MoTe2. *Nature* **622**, 63–68 (2023).
20. Chen, G. *et al.* Tunable correlated Chern insulator and ferromagnetism in a moiré superlattice. *Nature* **579**, 56–61 (2020).
21. Ding, J. *et al.* Electric-Field Switchable Chirality in Rhombohedral Graphene Chern Insulators Stabilized by Tungsten Diselenide. *Phys. Rev. X* **15**, 011052 (2025).
22. Xiang, H. *et al.* Continuously tunable anomalous Hall crystals in rhombohedral heptalayer graphene. Preprint at https://doi.org/10.48550/arXiv.2502.18031 (2025).
23. Wang, Z. *et al.* Electrical switching of Chern insulators in moiré rhombohedral heptalayer graphene. Preprint at https://doi.org/10.48550/arXiv.2503.00837 (2025).
24. Waters, D. *et al.* Chern Insulators at Integer and Fractional Filling in Moiré Pentalayer Graphene. *Phys. Rev. X* **15**, 011045 (2025).
25. Zeng, Y. *et al.* Thermodynamic evidence of fractional Chern insulator in moiré MoTe2. *Nature* **622**, 69–73 (2023).
26. Redekop, E. *et al.* Direct magnetic imaging of fractional Chern insulators in twisted MoTe2. *Nature* **635**, 584–589 (2024).
27. Ji, Z. *et al.* Local probe of bulk and edge states in a fractional Chern insulator. *Nature* **635**, 578–583 (2024).
28. Thompson, E. *et al.* Microscopic signatures of topology in twisted MoTe2. *Nat. Phys.* **21**, 1224–1230 (2025).
29. Liu, Y. *et al.* Imaging moiré flat bands and Wigner molecular crystals in twisted bilayer MoTe2. Preprint at https://doi.org/10.48550/arXiv.2406.19310 (2025).
30. Kang, K. *et al.* Evidence of the fractional quantum spin Hall effect in moiré MoTe2. *Nature* **628**, 522–526 (2024).
31. Park, H. *et al.* Ferromagnetism and topology of the higher flat band in a fractional Chern insulator. *Nat. Phys.* https://www.nature.com/articles/s41567-025-02804-0 (2025).
32. Xu, F. *et al.* Interplay between topology and correlations in the second moiré band of twisted bilayer MoTe2. *Nat. Phys.* **21**, 542–548 (2025).
33. Dong, J. *et al.* Anomalous Hall Crystals in Rhombohedral Multilayer Graphene. I. Interaction-Driven Chern Bands and Fractional Quantum Hall States at Zero Magnetic Field. *Phys. Rev. Lett.* **133**, 206503 (2024).
34. Zhou, B., Yang, H. & Zhang, Y.-H. Fractional Quantum Anomalous Hall Effect in Rhombohedral Multilayer Graphene in the Moiréless Limit. *Phys. Rev. Lett.* **133**, 206504 (2024).

35. Dong, Z., Patri, A. S. & Senthil, T. Theory of Quantum Anomalous Hall Phases in Pentalayer Rhombohedral Graphene Moiré Structures. *Phys. Rev. Lett.* **133**, 206502 (2024).
36. Guo, Z., Lu, X., Xie, B. & Liu, J. Fractional Chern insulator states in multilayer graphene moiré superlattices. *Phys. Rev. B* **110**, 075109 (2024).
37. Kwan, Y. H. *et al.* Moiré fractional Chern insulators. III. Hartree-Fock phase diagram, magic angle regime for Chern insulator states, role of moiré potential, and Goldstone gaps in rhombohedral graphene superlattices. *Phys. Rev. B* **112**, 075109 (2025).
38. Tan, T. & Devakul, T. Parent Berry Curvature and the Ideal Anomalous Hall Crystal. *Phys. Rev. X* **14**, 041040 (2024).
39. Crépel, V. & Cano, J. Efficient Prediction of Superlattice and Anomalous Miniband Topology from Quantum Geometry. *Phys. Rev. X* **15**, 011004 (2025).
40. Huang, K., Li, X., Das Sarma, S. & Zhang, F. Self-consistent theory of fractional quantum anomalous Hall states in rhombohedral graphene. *Phys. Rev. B* **110**, 115146 (2024).
41. Yu, J., Herzog-Arbeitman, J., Kwan, Y. H., Regnault, N. & Bernevig, B. A. Moiré fractional Chern insulators. IV. Fluctuation-driven collapse in multiband exact diagonalization calculations on rhombohedral graphene. *Phys. Rev. B* **112**, 075110 (2025).
42. Uzan, M. *et al.* hBN alignment orientation controls moiré strength in rhombohedral graphene. Preprint at https://doi.org/10.48550/arXiv.2507.20647 (2025).
43. Li, T. *et al.* Continuous Mott transition in semiconductor moiré superlattices. *Nature* **597**, 350–354 (2021).
44. Li, T. *et al.* Quantum anomalous Hall effect from intertwined moiré bands. *Nature* **600**, 641–646 (2021).
45. Li, C. *et al.* Tunable Chern Insulators in Moiré-Distant and Moiré-Proximal Rhombohedral Pentalayer Graphene. Preprint at https://doi.org/10.48550/arXiv.2505.01767 (2025).
46. Liu, K. Spontaneous broken-symmetry insulator and metals in tetralayer rhombohedral graphene. *Nat. Nanotechnol.* **19**, (2024).
47. Han, T. *et al.* Correlated insulator and Chern insulators in pentalayer rhombohedral-stacked graphene. *Nat. Nanotechnol.* **19**, 181–187 (2024).
48. Herzog-Arbeitman, J. *et al.* Moiré fractional Chern insulators. II. First-principles calculations and continuum models of rhombohedral graphene superlattices. *Phys. Rev. B* **109**, 205122 (2024).
49. Kolář, K., Waters, D., Folk, J., Yankowitz, M. & Lewandowski, C. Single-gate tracking behavior in flat-band multilayer graphene devices. Preprint at https://doi.org/10.48550/arXiv.2503.10749 (2025).
50. Dong, Z., Patri, A. S. & Senthil, T. Stability of anomalous Hall crystals in multilayer rhombohedral graphene. *Phys. Rev. B* **110**, 205130 (2024).
51. Soejima, T. *et al.* Anomalous Hall crystals in rhombohedral multilayer graphene. II. General mechanism and a minimal model. *Phys. Rev. B* **110**, 205124 (2024).
52. Bernevig, B. A. & Kwan, Y. H. 'Berry Trashcan' Model of Interacting Electrons in Rhombohedral Graphene. Preprint at https://doi.org/10.48550/arXiv.2503.09692 (2025).
53. Rezayi, E. H. & Simon, S. H. Breaking of Particle-Hole Symmetry by Landau Level Mixing in the ν = 5 / 2 Quantized Hall State. *Phys. Rev. Lett.* **106**, 116801 (2011).
54. Ring, P. & Schuck, P. *The Nuclear Many-Body Problem*. (Springer Science & Business Media, 2004).
55. Schuck, P. *et al.* Equation of Motion Method for strongly correlated Fermi systems and Extended RPA approaches. *Physics Reports* **929**, 1–84 (2021).

# Acknowledgements

We thank Zhida Song, Gengdong Zhou and Dmitri K. Efetov for the useful discussion. X.L. acknowledges support from the National Key R&D Program (Grant nos. 2022YFA1403500, 2022YFA1403502 and 2024YFA1409002) and the National Natural Science Foundation of China (Grant Nos. 12274006, 12404044 and 12141401). B.A.B. was supported by the Gordon and Betty Moore Foundation through Grant No. GBMF8685 towards the Princeton theory program, the Gordon and Betty Moore Foundation's EPiQS Initiative (Grant No. GBMF11070), the Office of Naval Research (ONR Grant No. N00014-20-1-2303), the Global Collaborative Network Grant at Princeton University, the Simons Investigator Grant No. 404513, the NSF-MERSEC (Grant No. MERSEC DMR 2011750), Simons Collaboration on New Frontiers in Superconductivity (SFI-MPS- NFS-00006741-01), and the Schmidt Foundation at the Princeton University, and the Princeton Catalyst Innitiative. N. R. Was supported partially also by a European Research Council (ERC) under the European Union's Horizon 2020 research and innovation program (Grant Agreement No. 101020833). The Flatiron Institute is a division of the Simons Foundation. K.W. and T.T. acknowledge support from the JSPS KAKENHI (Grant nos. 21H05233 and 23H02052) and World Premier International Research Center Initiative (WPI), MEXT, Japan.

# Author Contributions

X.L. and W.W. conceived and designed the experiments; J.X. Z.H. and W.W fabricated the devices and performed the transport measurement with help from others. Z.H., J.H.-A., Y.K., B.A.B. and X.L. analyzed the data; Y.K., J.H.-A., N.R. and B.A.B. performed the theoretical modeling; T.T. and K.W. contributed hBN substrates; Z.H., J.H.-A., Y.K., B.A.B. and X.L. wrote the paper with input from others.

# Competing interests

The authors declare no competing interests.

# Data and materials availability

All data are available in the main text or the supplementary materials.

# Supplementary Materials

Materials and Methods

Figs. S1 to S5

References

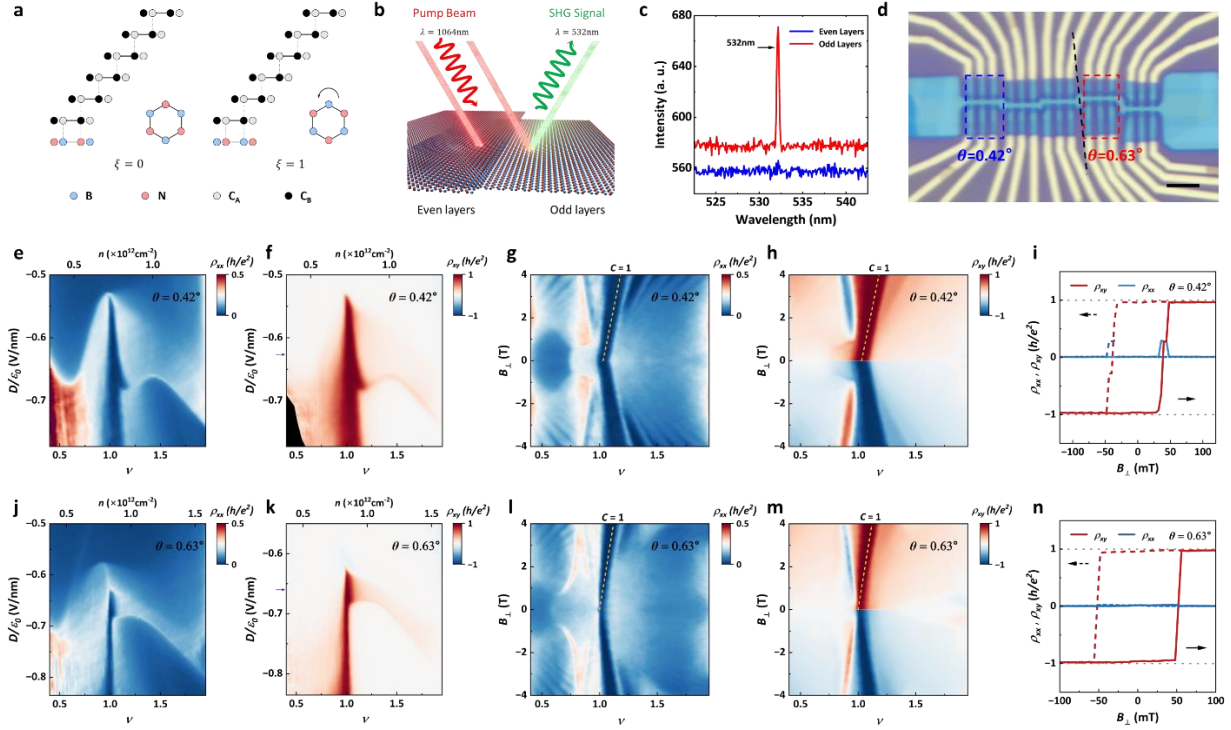

**Fig. 1 | Chern insulator in RHG/hBN with ξ = 0 and ξ = 1 configurations. a,** Schematic diagram showing ξ = 0 and ξ = 1 stacking configurations between RHG and hBN. **b,** Schematic diagram of SHG measurement. **c,** SHG signal measured at two sides of the monolayer step (blue and red points shown in Fig. S1b). The data of odd layers has been artificially offset by 20 a.u. to prevent overlap and facilitate clear comparison between datasets. **d,** Optical image of devices with ξ = 0 and ξ = 1 configurations. The monolayer step of hBN is indicated with a black dashed line. Blue (Red) dashed rectangle demarcates the location of device with $\theta = 0.42°$ ($\theta = 0.63°$). The two devices have different stacking configurations with ξ = 0 or ξ = 1. The scale bar in optical image represents 5 μm. **e,f,** Phase diagrams of symmetrized longitudinal resistivity $\rho_{xx}$ (**e**) and antisymmetrized Hall resistivity $\rho_{xy}$ (**f**) for D3 with $\theta = 0.42°$, as functions of moiré filling factor $v$ (carrier density $n$) and electric displacement field $D$ measured at $B_\perp = \pm 0.1$T and $T = 10$mK. Note that the $D < 0$ corresponds to the moiré-distant side. **g,h,** Landau fan diagrams of $\rho_{xx}$ (**g**) and $\rho_{xy}$ (**h**) for D3 at $D/\varepsilon_0 = 0.626$V/nm marked by the purple arrow in **f**. **i,** Magnetic hysteresis curves for D3 measured at suitable points ($n = 0.755 \times 10^{12}$cm$^{-2}$, $D/\varepsilon_0 = 0.648$V/nm). Dashed lines indicate the trajectories of gaps with Chern number $C = 1$. **j-n,** Comparative measurement results for D5 with $\theta = 0.63°$, similar to **e-i**. **l** and **m** were measured at $D/\varepsilon_0 = 0.660$V/nm, and **n** was measured at $n = 0.855 \times 10^{12}$cm$^{-2}$, $D/\varepsilon_0 = 0.668$V/nm. The data of $\rho_{xx}$ and $\rho_{xy}$ is symmetrized and antisymmetrized, respectively.

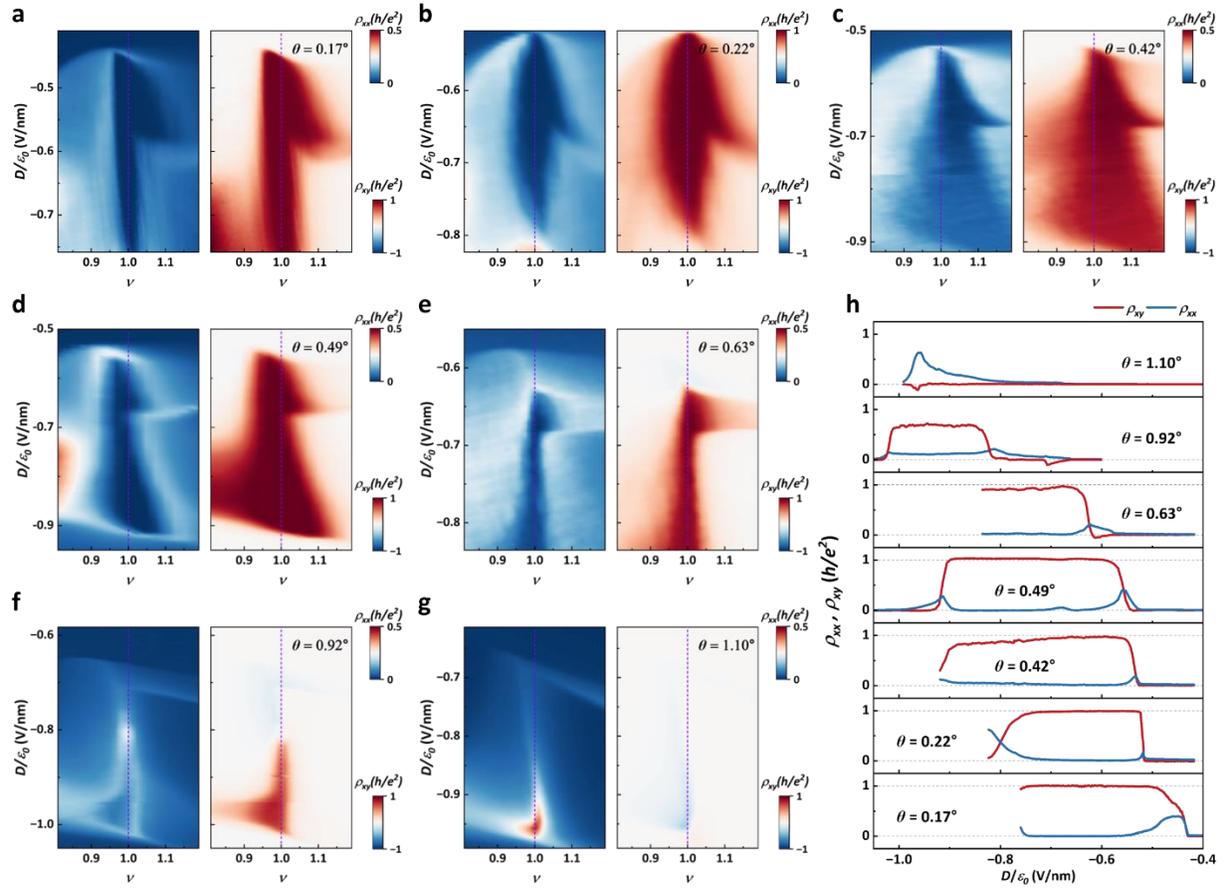

**Fig. 2 | Twist angle dependent band topology. a-g,** Phase diagrams of symmetrized $\rho_{xx}$ (left panels) and antisymmetrized $\rho_{xy}$ (right panels) versus $v$ and $D/\varepsilon_0$ around $v = 1$ at the moiré-distant side for D1-7 with different $\theta$, measured at $B_\perp = \pm 0.1$T and $T = 10$mK. **i,** $\rho_{xx}$ and $\rho_{xy}$ linecuts at $v = 1$ from **a-g** (indicated by purple dashed lines) as functions of $D/\varepsilon_0$, showing the $\theta$-dependence of $C = 1$ Chern insulator state. As moiré twist angle increases, the wide plateaus of $\rho_{xy}$ appear and disappear at stronger displacement field until the $C = 1$ state annihilates. At $\theta = 1.10°$, the $C = 1$ state vanishes completely at zero magnetic field.

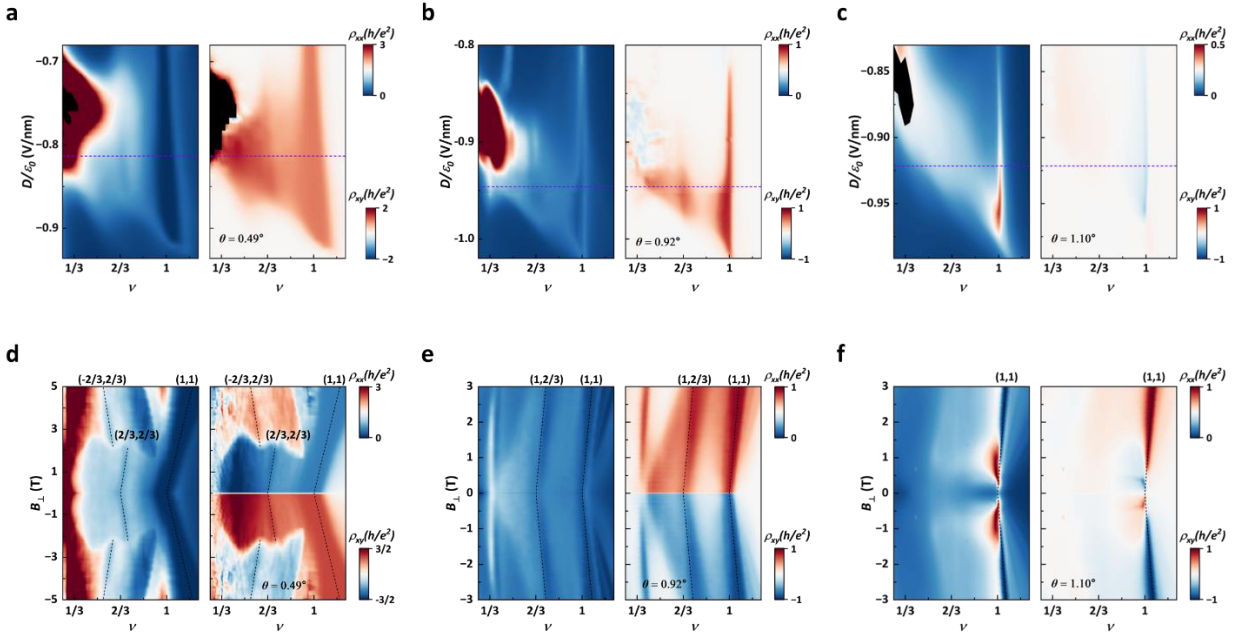

**Fig. 3 | Moiré dependent FCI state at $v = 2/3$. a-c,** Phase diagrams of symmetrized $\rho_{xx}$ (left panels) and antisymmetrized $\rho_{xy}$ (right panels) versus $v$ and $D/\varepsilon_0$ at the moiré-distant side for devices with $\theta = 0.49°$(**a**), $0.92°$(**b**), $1.10°$ (**c**), measured at $B_\perp = \pm 0.1$T and $T = 10$mK. The black regions are artifacts from the measurement because of bad contacts from high-resistance states. **d-f,** Landau fan diagrams of symmetrized $\rho_{xx}$ (left panels) and antisymmetrized $\rho_{xy}$ (right panels) versus $v$ and $B_\perp$ measured at $D/\varepsilon_0 = $ -0.813 V/nm for D4 (**d**), -0.946 V/nm for D6 (**e**), -0.922 V/nm for D7(**f**), marked by purple dashed lines in **a-c**. The black dashed lines are determined by Streda's formula and labeled as $(C, v)$, where $C$ is the coefficient in Streda's formula ($\partial n/\partial B_\perp = C \cdot e/h$) and $v_0$ is moiré filling factor at zero magnetic field. In D4 with $\theta = 0.49°$, a clear dip in $\rho_{xx}$ fully adhere to Streda's formula with $C = \pm 2/3$ and $v = 2/3$, accompanied by the sign reversal of $C$ at high magnetic fields. On the other hand, in the Landau fan diagram of D6 with $\theta = 0.92°$, the $v = 2/3$ state exhibits a Chern insulator with $C = 1$. In D7 with $\theta = 1.10°$, there is neither clear FCI nor ICI state observed at $v = 2/3$.

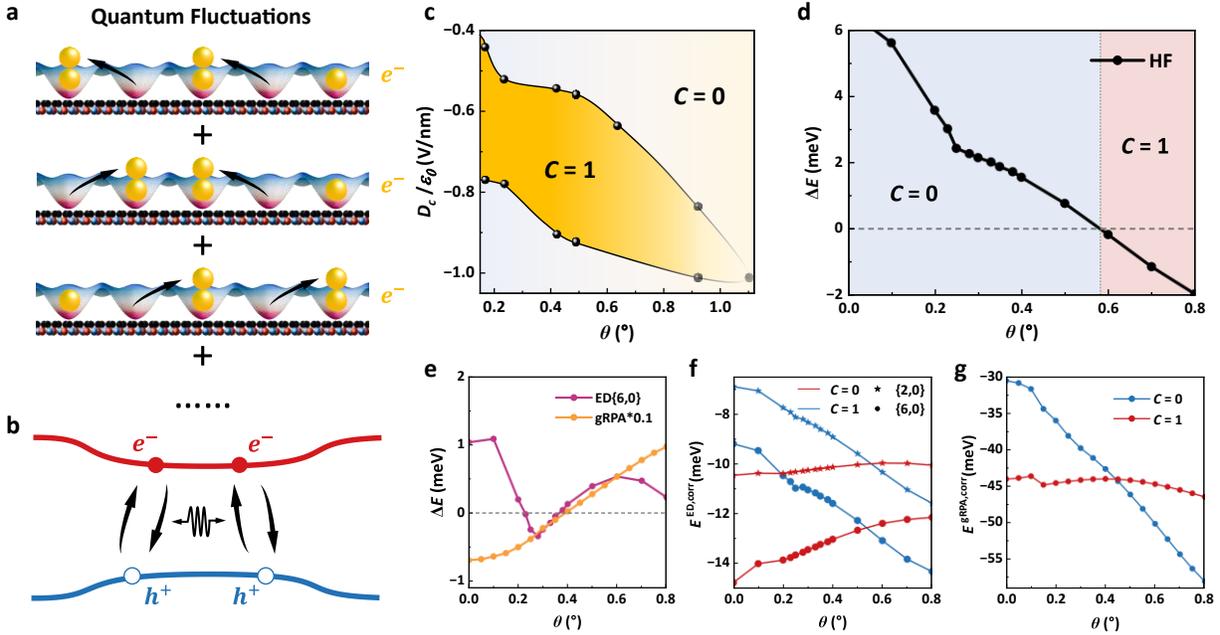

**Fig. 4 | The effect of quantum fluctuations on band topology. a,** Real-space schematic of quantum fluctuations (QFs) at $v = 1$. The lower parts depict the moiré superlattice and the corresponding moiré potential for simplicity. Yellow spheres represent filled electrons at $v = 1$. The QF ground state involves fluctuating particle-hole pairs (denoted by black arrows) on top of the HF ground state. Note that we do not assert any specific mechanism for QFs. **b,** Virtual processes in QFs. Red and blue curves schematically represent the occupied and unoccupied HF bands in the conduction subspace. Particle–hole pairs can momentarily appear (virtual excitation) and annihilate. Such processes contribute to corrections to the ground state energy beyond the mean-field. **c,** Schematic representation of $C = 1$ state versus $\theta$ and critical displacement field $D_c/\varepsilon_0$ at $v = 1$, showing the $\theta$-dependence of Chern insulator state. The black dots represent data points extracted from Fig. 2a-g. **d,e,** Total energy difference $\Delta E = E_{C=1} - E_{C=0}$ between the $C = 0,1$ states, computed using HF, ED and the gRPA. Calculations are performed on a 21-site lattice at $v = 1$ (see Methods). For HF, The blue-shaded (red-shaded) region represents the portion where the mean-field calculation yields a ground state with Chern number $C = 0$ ($C = 1$). For ED, $\{N,0\}$ indicates that up to $N$ particles are permitted to fluctuate to the lowest unoccupied HF band. Note that the gRPA energy difference has been scaled down by a factor of 0.1 to visually emphasize the overall trend, considering its significantly larger (overestimated) correlation energies. **f,g,** Correlation energy relative to the HF energy of the $C = 0,1$ states, computed using ED ($E^{\text{ED,corr}} = E^{\text{ED}} - E^{\text{HF}}$) and gRPA ($E^{\text{gRPA,corr}} = E^{\text{gRPA}} - E^{\text{HF}}$) respectively. Across all parameters studied, in ED and in gRPA, the correlation energies of total energy difference $\Delta E^{\text{corr}}$ show a $\theta$-dependence trend opposite to the HF energy difference $\Delta E^{\text{HF}}$, and are qualitatively consistent with the topological phase boundary measured in transport.